\documentstyle[twocolumn,aps,prl]{revtex}

\begin{document}
\draft
\title{ Breakup of a Dimer: A New Approach to Localization Transition }
\author{ Ignacio G\'omez\cite{paddress} and Indubala I. Satija\cite{email}}
\address{
 Department of Physics, George Mason University,
 Fairfax, VA 22030}
\date{\today}
\maketitle
\begin{abstract}
Within the framework of tight binding models,
aperiodic systems are mapped to a renormalized lattice with a dimer defect. 
In models exhibiting metal-insulator transition, the dimer acts like a resonant cavity
and explains the existence of the ballistic transport in the system.
The localization in the model can be attributed to the vanishing
of the coupling between the two sites of the dimer.
Our approach unifies Anderson transition and resonance transition
and provides a new formulation to understand localization and its
absence in aperiodic systems.

\end{abstract}

\narrowtext

The existence of metal-insulator transitions due to quantum interference in correlated
disordered systems is a fascinating subject. The origin of this resonant transition
can be understood from the textbook examples of the quantum mechanics of a barrier
in the continuum problem and from a simple dimer model\cite{dimer}
in the case of a discrete lattice.
The resonance transition
where the metallic phase with ballistic transport is
due to zero reflectance at the two sites of the dimer provides a simple mechanism
to understand localization and its absence in systems with short range correlation.
Recently, this transition was verified in experiments 
in superlattices\cite{Bellini}.

In this letter, we present a universal approach to understand
metal-insulator transitions by unifying Anderson transition with
the resonance transition.
Our formulation is applicable for any $aperiodic$ system 
with reflection symmetry
described by a nearest-neighbor tight binding model (TBM).
In this paper we will confine ourselves to sinusoidal potentials
which are further modulated by a Gaussian profile.
The purpose of the Gaussian modulation is two-fold:
as explained below, it provides a useful means to motivate and illustrate
our ideas even though our results and conclusions are valid in the limit
where the width of the Gaussian goes to $\infty$.
Secondly, it facilitates the study of
the pure Gaussian
systems that
have been the subject of recent studies\cite{gauss} due to its possible application
as efficient energy band-pass filters
in semiconducting superlattices. Finally, the case of sinusoidal potential
further modulated by a Gaussian profile includes the famous Harper
equation exhibiting Anderson localization as a limiting case.\cite{Harper}

The model system under consideration here is a TBM describing an 
eigenvalue problem with energy $E$,
\begin{equation}
\psi_{m+1} + \psi_{m-1} - E_m \psi_m = 0.
\end{equation}
The $E_m=E-\epsilon_m$ is the diagonal term containing the aperiodic onsite
energy $\epsilon_m$.
This model also describes
the Schrodinger equation 
for an array of
$\delta$-function Kronig-Penny potential barriers, 
\begin{equation}
-\frac{\hbar^2} {2M} \psi^{''}(x)+ \sum \epsilon_m \delta(x-ma) \psi(x)=E \psi(x).
\label{kp}
\end{equation}
This is due to the fact that the Poincare map
\cite{Jose} associated with the model is a TBM with
$E_m = \epsilon_m \frac{sin(K)}{K}-cos(K) $. 
Here $K$ is the Bloch vector related to the energy as $E=\hbar^2 K^2/2M$.

We choose $\epsilon_m$ to be sinusoidal potential which is
further modulated by a Gaussian,
\begin{equation}
\epsilon_m = 2 \lambda \cos [ 2 \pi \mu (m-m_0)] \exp [-\frac{(m-m_0)^2}
{2\sigma^2}] .
\end{equation}
Here $\lambda$ is the strength
of the potential, $\sigma$ is the width of the Gaussian profile and $\mu$
is an irrational number chosen to be the inverse golden-mean
$\mu=\frac{\sqrt{5}-1}{2}$.
In this paper, we will describe our results for the pure Gaussian model ($\mu=0$),
the Harper equation ($\sigma \to \infty$) and also for the $\delta$-function
barriers Eq. (\ref{kp})
The last case closely resembles the recent study of Gaussian
modulated Kronig-Penny 
model\cite{gauss} 
and is also currently under experimental investigation in superlattices.\cite{private}. 

The basic idea underlying our approach can be understood
for any localized defect
with finite spatial extent. For aperiodic systems such as Harper equation
where the aperiodicity exists throughout
the lattice, we introduce Gaussian modulation so that the system
can be viewed as a lattice with a localized defect.
However, as shown below, 
the $\sigma \to \infty$ limit is well defined and therefore Gaussian modulation
,although not needed, is useful in understanding the decimation scheme
described below. 

We envision perfectly transmitting phase in all aperiodic systems to be described by
Bloch states on some "renormalized" lattice. In aperiodic systems where the defects 
are spatially confined to only some parts of the lattice, Bloch wave  amplitudes are
attenuated or amplified due to various quantum interferences
only in the neighborhood of these defects. ( See figure ~1)
We eliminate such sites from the model
by decimating them. The resulting renormalized model will have solutions that are
Bloch waves at all sites in the metallic phase. 
We would like to emphasize that we decimate $all$ sites which is in contrast to
previous use of decimation for random systems where every other
site is decimated. 
However, the basic idea of decimating all sites is similar
in spirit to that of Fibonacci decimation\cite{KSRG} used in quasiperiodic systems
where one eliminates 
all but Fibonacci sites so that the renormalized lattice may exhibit translational
invariance in Fibonacci space.

Figure ~2 outlines the decimation process.
We begin by eliminating the central site $m_0$ of the symmetric defect. 
This leads to a renormalized model consisting of a dimer with coupling between the two
sites as $\bar{\gamma}_1=1/E_0$ and the onsite energy
$ \bar{E}_1 = E_1 - \bar{\gamma}_1$.
We now begin the
iterative process by decimating
this dimer. At the $n^{th}$ step, we obtain a new lattice with renormalized dimer
with onsite energies denoted as $\bar{E}_{n+1}$ and the renormalized coupling between the 2-sites
of the dimer as $\bar{\gamma}_{n+1}$.
This iterative decimation scheme results in a two-dimensional $driven$ map
for $\bar{\gamma}-\bar{E}$ where the $E_n$ containing diagonal disorder
of the bare model provides the driving term,
\begin{eqnarray}
\bar{\gamma}_{n+1} &=& \bar{\gamma}_{n} \over { {\bar{E}_n}^2-{\bar{\gamma}_{n}}^2 }
\nonumber\\
\bar{E}_{n+1} &= &E_{n+1} - (1+{\bar{\gamma}}_{n+1} \bar{\gamma}_{n})/ \bar{E}_n.
\label{2d}
\end{eqnarray}

It should be noted that the parameters of the model are included in $E_{n+1}$.
This 2-dimensional map contains
the complexity of
various interference effects within the Gaussian profile, manifesting itself
in the energy dependent coupling and onsite potential.
We would like to emphasize again that the map has a well defined limit for
$\sigma$ equal to $\infty$. Therefore, Gaussian profile is not necessary in
obtaining the mapping of the aperiodic model to the dimer model. The dynamics of
the map does depend upon $\sigma$: 
the finite value of $\sigma$ provides damping in the map
with the consequence that the renormalization
group (RG) flow settles on attractors while, in the limit of
$\sigma \to \infty$, there are no attractors in the map.

The usefulness of this map emerges from the fact that the changes in
the parameters $\lambda$
are reflected in the significant changes in the trajectories of the map
for both finite as well as for the infinite value of $\sigma$.
Figure ~3 shows how the metal-insulator transition in the pure Gaussian case
manifests itself in terms of the variation in the RG attractor as $\lambda$ is varied.
In the subcritical phase, where the lattice
has a transmission coefficient $T$ equal to unity,
a symmetric period-2 limit cycle describes the RG flow for the renormalized
coupling $\bar{\gamma}$ while the dynamics of 
$\bar{E}$ is governed by a fixed point. 
Away from the transition point, the RG attractor
exhibits very regular oscillatory pattern consisting of symmetric lobes and divergences
which is periodic in $\lambda$ with periodicity
equal to $\frac{1}{\sigma}$. 
The fact that the total number of lobes or divergences 
is equal to $\sigma$ suggests that each lobe owes its existence to
a particular site within the Gaussian profile where the sites near the center contributing
at smaller values of $\lambda$.
However, the quantitative understanding of almost equally spaced lobes and its physical
significance remains elluded to us at present. 

As we approach the transition point, the symmetric
2--cycle of $\bar{\gamma}$ looses its symmetry,
degenerates to a fixed point (seen in the figure~3 where the lobes cross)
and then
continues as an asymmetrical period-2 attractor. On the other hand,
the fixed point describing the $\bar{E}$ values becomes a 2--cycle.
At the transition, $\bar{\gamma}$ approaches zero 
with a power-law decay resulting in a $broken$ $dimer$. 
The localized phase is characterized by an exponential vanishing of the
the coupling
with a characteristic length which is found to be related to the localization length $\xi$,
of the localized wave function, $\bar{\gamma}_n \approx exp-(2 n/\xi)$.
It turns out that in the localized phase, $\bar{E}$
continues to be described by a period--2 fixed point exhibiting divergences 
where the spacing between two successive divergences increases
as $\lambda$ increases. 

The localization transition discussed above is a resonant transition where the metallic
phase is described by
Bloch wave solutions 
which undergo real phase shifts as they encounter the dimer defect.
In the localized phase, these phase shifts become imaginary.
Imposing this condition on the solutions of the renormalized system
determines the condition for the perfect transmission.
We express this condition in terms of a function $f$ defined as 
\begin{equation}
f_n(\lambda, \sigma, E) = 1-{\bar{\gamma}_{n}}^2+{\bar{E}_n}({\bar{E}_n}-E_n).
\end{equation}
Figure ~4 shows the variation in $f_n$ and the transmission coefficient $T$
for the Kronig-Penny model Eq. (\ref{kp}) as the energy $E$
is varied. We see that the condition for the vanishing of this function coincides with 
the condition for perfect transmission. 
We would like to point out that the transmission coefficient was calculated
in a rather simple way by using the renormalized dimer \cite{Trans}.
This requires multiplying only two transfer matrices in contrast to usual
calculations where one needs to multiply all transfer matrices on an
aperiodic lattice. 

The existence of a band of conducting states in Gaussian modulated lattices as discussed
above ( which is consistent with the previous related study\cite{gauss}) 
is a very desirable feature of a
lattice making it a useful filter. Attempts are underway\cite{private}
to make such filters using superlattices.
This should be contrasted with the  dimer-type defects where the metal-insulator
transition is obtained by fine tuning the parameter to obtain the resonant energy
close to the Fermi energy.
We would like to point out that the metal-insulator transition and a band of
perfectly transmitting states in Gaussian system can also be
understood by heuristic arguments of asymptotic property of "constantcy"
of the potential. Due to the finite width of the potential, the model
has a constant potential asymptotically. This property is crucial in
determining the $E$, which is the global property \cite{Sarma}. This argument
leads to extended states if $E > 2-2 \lambda$ and localized states
for $E < 2-2 \lambda$. This condition is found to be true in our numerical
simulations for large values of $\sigma$.

We next discuss the case of the pure Harper equation obtained in the limit
$\sigma \to \infty$ in the driving term for the two-dimensional map.
The sub-critical phase is not an attractor. 
As $\lambda$ increases, the RG trajectories become more and more complex ( see figure ~5)
\cite{www}
and eventually collapse to a vertical line corresponding to $\bar{\gamma} \to 0$
at the onset to localization transition.
The localized phase is again characterized by an
exponentially decaying coupling of the dimer with
length scale which is equal to half the localization of the Harper equation.
The metallic phase in Harper equation is thus described in terms of resonace
due to the dimer and the Anderson localization is due to the breaking of this
dimer. 


Recently, the existence of extended states
in supercritical Frenkel-Konterova(FK) model and in Fibonacci lattices was shown to be due to 
dimer-type correlations\cite{FK} using decimation schemes that were model
dependent. The novel aspect of the work described here is the universal
nature of our approach:
the 2--dimensional map Eq. (\ref{2d}) can be used
to study localization or its absence in systems with short range correlations
such as dimer defects, FK and Fibonacci models with long range correlations
as well as in Harper equation. 
Our most important result is the picture of the localized phase 
as a phase with the broken dimer. 
Although the present framework is developed only
for the TBM type systems, it is possible to extend our method to study localization
in 2-dimensional
models as well as for systems with long range interaction.
The new methodology developed here will have applications
in many other areas including
dynamical localization in kicked rotors\cite{QC}
as well as the transition to strange nonchaotic attractors(SNAs)
in quasiperiodically driven maps \cite{KSproc,ft}.

The research of IIS is supported by National Science
Foundation Grant No. DMR~097535. 
IGC would like to thank for the hospitality
during his visit to George Mason University.
We would like to thank Bala Sundaram for his useful comments on this paper.

\begin{figure}
\caption{ The wave function for
sinusoidally modulated Gaussian case. We have Bloch wave with $K=1/2$ outside the
Gaussian profile. The points in the wave function are not joined for clarity. Here $\lambda=0.5$ and $\sigma=1000. $}
\label{fig1}
\end{figure}

\begin{figure}
\caption{ The Gaussian modulated
lattice and the iterative decimation process.}  
\label{fig2}
\end{figure}

\begin{figure}
\caption{ The variation in the asymptotic values of the
renormalized coupling $\bar{\gamma}_n$ and $\bar{E}_n$ for $\sigma=100$. There are
few isolated points in the localized phase near the transition where $\bar{\gamma}_n$
may diverge. Note that $tanh$ is used to display the behavior at $\pm \infty$.} 
\label{fig3}
\end{figure}

\begin{figure}
\caption{ The
band of resonant energies corresponding to perfect transmission for
$\delta$-function potential Kronig-Penny model.
Here $\lambda=10$ and $\sigma=100$.} 
\label{fig4}
\end{figure}

\begin{figure}
\caption{RG trajectory for $\lambda=0.992$
for $E=0$ case for the Harper equation.}
\label{fig5}
\end{figure}


\begin{references}
\bibitem{ft}It is interesting to note that 
in the localized phase the RG equation for $\bar{E}_n$ is identical
to the original Harper equation with $\bar{E}_n = \psi_{n+1}/\psi_n$.
This map has been studied recently as a model for strange nonchaotic attractors
(SNA) in quasiperiodically driven maps.\cite{KSproc}
\end{references}
\end{document}